\documentclass[traditabstract]{aa} 
 
\usepackage{graphicx}
\usepackage{txfonts}
\usepackage{natbib}
\bibpunct{(}{)}{;}{a}{}{,}

\begin{document}

   \title{Blue supergiant progenitor models of Type II supernovae
         }

   \author{D. Vanbeveren
           \and N. Mennekens
           \and W. Van Rensbergen
           \and C. De Loore
          }

   \institute{Astrophysical Institute, Vrije Universiteit Brussel, Pleinlaan 2, 1050 Brussels, Belgium\\
              \email{dvbevere@vub.ac.be}
             }

   \date{Received }

  \abstract
   {In the present paper we show that within all the uncertainties that govern the process of Roche lobe overflow in Case Br type massive binaries, it can not be excluded that a significant fraction of them merge and become single stars. We demonstrate that at least some of them will spend most of their core helium burning phase as hydrogen rich blue stars, populating the massive blue supergiant region and/or the massive Be type star population. The evolutionary simulations let us suspect that these mergers will explode as luminous hydrogen rich stars and it is tempting to link them to at least some super luminous supernovae.}
{} {} {} {}
   \keywords{binaries: close -- binaries: evolution
            }
            
   \titlerunning{Blue supergiant progenitor models of Type II supernovae}
   
   \authorrunning{D. Vanbeveren et al.}

   \maketitle

\section{Introduction}

Most type II supernova progenitors are red supergiants but SN 1987A in the Large Magellanic Cloud provided strong evidence that the progenitor of a Type II supernova can also be a massive hydrogen rich blue supergiant with a mass $\sim$ 20 M$_{\odot}$ (Arnett et al., 1989).

Since the discovery in 1999 of the first superluminous supernovae (SLSN) (e.g., SN1999as, Knop et al., 1999; and SN 1999bd, Nugent et al., 1999) recent surveys have detected numerous SLSN events: they are most likely associated with the death of the most massive stars (for a review, see Gal-Yam, 2012). About 30-40\% have strong hydrogen lines in their spectra (they therefore get the SLSN-II subtype classification) indicating that the explosions happen in a thick hydrogen envelope, i.e., SLSN-II are explosions of the most massive stars that retained their hydrogen envelope until they exploded. Furthermore, Smith et al. (2007, 2008) presented evidence that the progenitors of SN 2006gy and SN 2006tf lost a large amount of mass prior to explosion suggesting that these progenitors may have been similar to massive luminous blue variables (LBVs).

To explain the high energy of SLSN-II explosions mainly three models have been proposed in literature. A pair-instability explosion is a first possibility (Woosley et al., 2007) however, the expected late time radioactive decay rate of $^{56}$Co is not observed. Furthermore, the $^{56}$Ni in SLSN-II estimated from observations does not agree with the large amount expected from theoretical pair-instability supernova models (Gal-Yam, 2012). An alternative is the spin-down of nascent rapidly rotating neutron stars with strong magnetic fields (magnetars) (Woosley, 2010; Kasen et al., 2010) and the third model is related to the collapse of massive stars into massive black holes where the energy of the supernova is produced by the rapid accretion of matter onto the black hole (MacFadyen and Woosley, 1999).

In the present paper we discuss evolutionary scenarios where Type II supernova progenitors are hydrogen rich blue supergiants. In section 2 we summarize the scenarios that have been proposed in the past. Section 3 deals with an alternative. The consequence of this alternative for overall massive star population synthesis is then considered in section 4.

\section{Evolution scenarios of blue progenitors of Type II supernovae: a review.}

\subsection{Single star progenitor models of SN 1987A}

Langer (1991) investigated the effect of semiconvection on the evolution of massive single stars with initial mass between 7 M$_{\odot}$ and 32 M$_{\odot}$ and with metallicity Z = 0.005.  By fine tuning the parameters governing the process of semiconvection, he concluded that single star models where convective core overshooting is ignored predict the existence of blue supergiant progenitors of Type II supernova.

Most of the single star evolutionary codes use the de Jager et al. (1988) prescription as a standard to compute the stellar wind mass loss rate during the red supergiant (RSG) phase of a massive star. However, as argued in Vanbeveren et al. (1998a, b) this standard is subject to large uncertainties and the real rates may be much higher. Even more, stellar evolutionary models with higher RSG rates explain the observed Wolf-Rayet star population much better (Vanbeveren et al., 1998a, b, 2007; Sander et al., 2012). In a recent paper, Georgy (2012) applied higher RSG rates in the Geneva single star evolutionary code and predicted in this way yellow and blue progenitors of Type-II supernovae.

\subsection{Binary progenitor models of SN 1987A}

It is known for quite some time now (see the references below) that a process which increases the fractional mass of the helium core, favours redward evolution during hydrogen shell/core helium burning (e.g. convective core overshooting during core hydrogen/helium burning). A process which on the contrary reduces this fractional mass will tend to keep a hydrogen shell/core helium burning star in the blue supergiant region of the Hertzsprung-Russell (HR) diagram. If, furthermore, such a process is able to form a hydrogen profile outside the helium core assuring a large fuel supply for the hydrogen burning shell, we may expect this star to remain blue up to the SN explosion. Close binary evolution provides us with two such processes: accretion onto a hydrogen shell burning star and the merger of both components at the moment that at least one of them is a hydrogen shell burning star.

The evolutionary behaviour of stars that accrete mass while being in the hydrogen shell burning stage, has been studied from phenomenological point of view by Hellings (1983, 1984). Full binary computations have been presented by Podsiadlowski et al. (1990), Podsiadlowski et al. (1992), De Loore and Vanbeveren (1992). They allow to conclude that mass gainers in Case B/C binaries may end their life as blue supergiants but the initial mass ratio of these binaries must be close to 1 making the occurrence frequency very low. Braun and Langer (1995) reconsidered the study of Hellings mentioned above, by using their favorite model to describe semiconvection. With this model the occurrence frequency becomes much larger. A more recent study by Claeys et al. (2011) essentially confirms the previous work although they show that also with the binary model fine-tuning is required.

A second binary scenario was presented by Podsiadlowski et al. (1992): the merger of the two components when the system evolves through a common envelope phase. They argued that this is expected to occur in late case B/case C binaries with mass ratio significantly smaller than 1. Detailed computations were performed for a 16 M$_{\odot}$  red giant that merged with a 3 M$_{\odot}$ main-sequence star. The authors concluded that the merger most likely ends its life as a blue supergiant. 

\subsection{Progenitor models of SLSN-II}

SLSN-II are very luminous events and difficult to reconcile with classical Type-II supernovae. The high luminosity led scientists to speculate that SLSN-II are pair creation supernova (PCSN) and thus that the progenitor stars are very massive stars with an initial mass larger than 100 M$_{\odot}$. Evolutionary calculations of Langer et al. (2007) illustrate that within the uncertainties of the stellar wind mass loss rate, very massive stars with an initial metallicity Z $\le$ Z$_{\odot}$/3 may retain their hydrogen-rich envelope and end their life with a PCSN. However, PCSN leave a unique chemical imprint, which would be observable in extremely metal-poor halo stars (Heger and Woosley, 2002) and this is not observed (Umeda and Nomoto, 2005). If PCSN did not take place in the first metal poor stellar generation, it is highly questionable that they would happen in stellar populations with nearly solar metallicity. Furthermore, as mentioned already in the introduction, SLSN-II observed at late times do not follow the $^{56}$Co radiative decay rate that is expected when the engine is a PCSN, whereas the amount of $^{56}$Ni observed in some of these supernovae is significantly less than the values resulting from theoretical models of pair-instability supernovae. 

\section{Massive Case Br binary mergers as blue supergiant progenitors of Type-II supernovae}

The detailed study of the evolution of massive binaries started in the sixties and since then it has been the subject of numerous papers. An idea how our massive binary knowledge evolved in time can be obtained by considering the following reviews (and references therein): Paczynski (1971), Van den Heuvel (1993), Vanbeveren (1993, 2009), Vanbeveren et al. (1998a), Langer (2012).

One of the uncertainties of (massive) close binary evolutionary scenarios is related to the process of Roche lobe overflow in general, in Case Br\footnote{Case Br binaries are binaries with an orbital period such that the Roche lobe overflow starts when the primary (the mass loser) is a hydrogen shell burning star with a hydrogen rich envelope that is mainly in radiative equilibrium.} binaries  in particular. The main question is whether Case Br Roche lobe overflow is conservative or not, and, if matter leaves the binary, what is the driving mechanism. More than 3 decades ago, our group introduced the parameter $\beta$ (what's in a name) as the relative amount of mass lost by the primary due to Roche lobe overflow that is accreted by the secondary (Vanbeveren et al., 1979). Obviously, $0 \le \beta \le 1$ but despite many efforts we think that it is fair to state that $\beta$ is still largely unknown. We will come back to this in section 4 where we will also argue that, accounting for the $\beta$-uncertainty, it cannot be excluded that the Roche lobe overflow in many Case Br massive binaries leads to the merger of both components.

The merger process is poorly understood and it is unclear how matter of the two components will mix together. However, in the case of Case Br mergers, a reasonable model may look like the one below. At the onset of the merging process in most of the Case Br binaries the lower mass companion (the secondary) is a hydrogen rich core hydrogen burning star\footnote{Only when the mass ratio of the binary is very close to 1, also the secondary may have a hydrogen-free core but the frequency of these type of binaries is expected to be very low (De Loore and Vanbeveren, 1992).}. The most massive component (the primary) however has a hydrogen-free core in radiative equilibrium, surrounded by a convective hydrogen burning shell. The core has a higher mean molecular weight and lower entropy than the surrounding hydrogen rich layers and it can be expected that during the merger the hydrogen rich mass of the secondary will mix mainly with these surrounding layers. This process therefore leads to the formation of a star with an undermassive helium core and, as stated in section 2.2, it can be expected that it will remain a blue supergiant during a considerable part of the core helium burning phase and perhaps they may explode while being blue. To illustrate we use the Brussels stellar evolutionary code\footnote{At present our binary evolution code is a twin code that follows the evolution of both components simultaneously (the code has been described in detail in Vanbeveren et al., 1998a, b; see also Vanbeveren et al., 2012). The opacities are taken from Iglesias et al. (1992), the nuclear reaction rates from Fowler et al. (1975). Semi-convection is treated according to the criterion of Schwarzschild and Harm (1958) and convective core overshooting is included as described by Schaller et al. (1992).} to calculate the evolution of a 30 + 20 M$_{\odot}$ Case Br binary with initial chemical composition (X, Y, Z) = (0.7, 0.28, 0.02). Immediately after the onset of Roche lobe overflow we assume that both components merge. The merging process is simulated as if the 20 M$_{\odot}$ secondary is accreted on the primary and we use an (ad hoc) accretion rate of $4 \cdot 10^{-2}$ M$_{\odot}$/yr. The merging process therefore takes about 500 yrs. We investigated the effect of different accretion rates and it can be concluded that as long as the accretion rate is such that the merging timescale remains smaller than the thermal timescale of the primary, the overall results are very similar to those presented here. The accreted layers are instantaneously mixed with the outer layers of the primary. To mix the layers we use the thermohaline mixing routine in our code. Thermohaline mixing is implemented in our code according to the description of Kippenhahn et al. (1980). After the merger process, the evolution of the star is further followed till the end of core helium burning.

\begin{table*}
\centering
\caption{Summary of the evolution of a 30+20 M$_{\odot}$ merger. Time is in $10^6$ yrs, the labels correspond to the labels in Figures 1-3, in the last column we first give the mass of the convective hydrogen burning core, then we give the mass of the hydrogen exhausted helium core, the last one is the mass of the CO core.}
\begin{tabular}{c c c c c c c c}
\hline
Label & Time & Mass (M$_{\odot}$) & Log T$_\mathrm{eff}$ & Log L/L$_{\odot}$ & X$_\mathrm{c}$ & Y$_\mathrm{c}$ & M$_\mathrm{core}$ (M$_{\odot}$) \\
\hline
 & 0 & 30 & 4.59 & 5.03 & 0.7 & 0.28 & 18 \\
A & 6.88 & 27.4 & 4.23 & 5.35 & 0 & 0.98 & M$_\mathrm{He}$ = 10 \\
B & 6.88 & 47.4 & 3.94 & 5.71 & 0 & 0.98 & M$_\mathrm{He}$ = 10 \\
C & 6.89 & 47.3 & 4.29 & 5.66 & 0 & 0.98 & M$_\mathrm{He}$ = 10 \\
D & 7.35 & 45.5 & 4.1 & 5.6 & 0 & 0 & M$_\mathrm{CO}$ = 9.2 \\
\hline
\end{tabular}
\end{table*}

\begin{figure*}[t]
\centering
\vspace{2cm}
   \includegraphics[width=14cm]{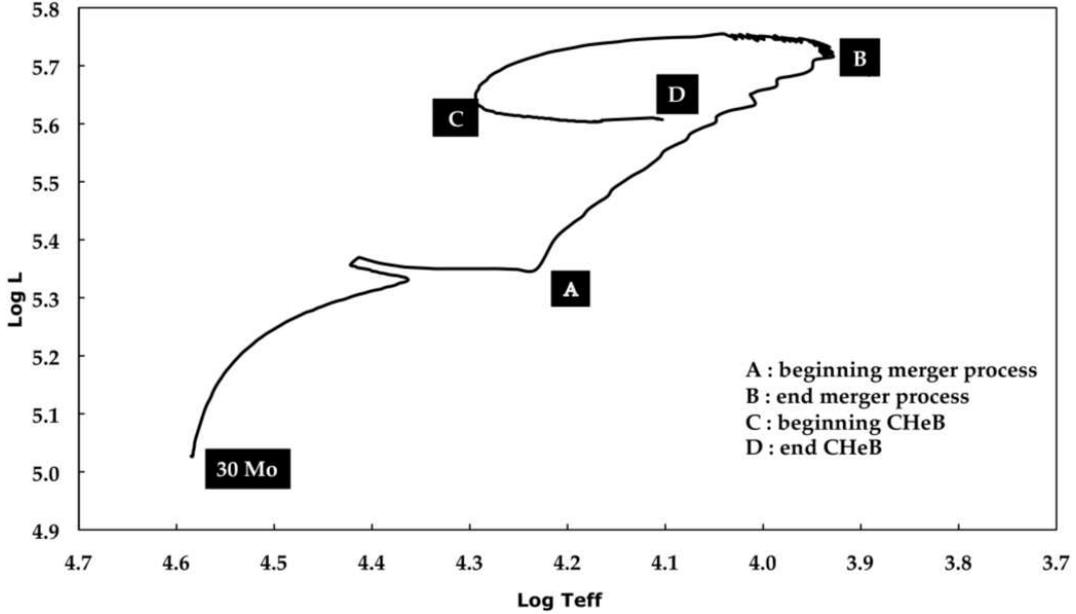}
     \caption{The evolutionary track of a 30 M$_{\odot}$ star in a Case Br binary that merges with its 20 M$_{\odot}$ companion.}
\end{figure*}

The evolution of the 30 M$_{\odot}$ star in the HR-diagram is shown in Figure 1 and summarized in table 1. Before the onset of Roche lobe overflow (point A) both stars are subject to stellar wind mass loss\footnote{To calculate the stellar wind mass loss rate of massive stars in the blue part of the HR-diagram we use in our evolutionary code the formalism of De Jager et al. (1988) as long as the atmospherical hydrogen abundance X$_\mathrm{atm} > 0.3$. Stars with X$_\mathrm{atm} \le 0.3$ are considered as Wolf-Rayet (WR) stars however, anticipating, the merger stars discussed in the present paper never become WR stars. Since 1988, various other stellar wind mass loss formalisms have been proposed (e.g., Vink et al., 2000; Pauldrach et al., 2012) but they all predict very similar results as far as the simulations of the present paper are concerned.}. The merger process starts at point A and lasts till point B. The merger is out of thermal equilibrium and is very extended. When the merging process stops (point B), the star quickly restores its equilibrium and evolves towards point C. At point C, helium starts burning in the core and point D marks the end of core helium burning. During most of the core helium burning phase, the star resembles a hydrogen rich blue supergiant with X$_\mathrm{atm} > 0.3$ and it therefore loses mass as predicted by the stellar wind mass loss rate formalism of De Jager et al. (1988) (see also footnote 4).  There are two reasons why the post-merger remains in the blue part of the HR-diagram. Firstly, the star has a small helium core with respect to its total mass (its total mass $\sim$46 M$_{\odot}$ but it has a He core of a 30 M$_{\odot}$ star). Secondly, the mixing of the secondary mass layers with the outer layers of the primary implies the formation of a hydrogen profile outside the helium core assuring a large fuel supply for the hydrogen burning shell and therefore the latter only slowly moves outwards. Interestingly, the post-merger blue supergiant has an atmospherical helium abundance Y $\sim$ 0.35 and is significantly nitrogen enriched (N/N$_{0}$ = 3-5).

\begin{table*}
\centering
\caption{Similar as Table 1 but for the 15+9 M$_{\odot}$ merger.}
\begin{tabular}{c c c c c c c c}
\hline
Label & Time & Mass (M$_{\odot}$) & Log T$_\mathrm{eff}$ & Log L/L$_{\odot}$ & X$_\mathrm{c}$ & Y$_\mathrm{c}$ & M$_\mathrm{core}$ (M$_{\odot}$) \\
\hline
 & 0 & 15 & 4.48 & 4.26 & 0.7 & 0.28 & 6.8 \\
A & 14.65 & 15 & 4.28 & 4.73 & 0 & 0.98 & M$_\mathrm{He}$ = 3.6 \\
B & 14.65 & 24 & 4.25 & 5.2 & 0 & 0.98 & M$_\mathrm{He}$ = 3.6 \\
C & 14.66 & 24 & 4.46 & 5.14 & 0 & 0.98 & M$_\mathrm{He}$ = 3.6 \\
D & 15.61 & 24 & 4.19 & 5.17 & 0 & 0 & M$_\mathrm{CO}$ = 3.2 \\
\hline
\end{tabular}
\end{table*}

\begin{table*}
\centering
\caption{Similar as Table 1 but for the 9+6.7 M$_{\odot}$ merger.}
\begin{tabular}{c c c c c c c c}
\hline
Label & Time & Mass (M$_{\odot}$) & Log T$_\mathrm{eff}$ & Log L/L$_{\odot}$ & X$_\mathrm{c}$ & Y$_\mathrm{c}$ & M$_\mathrm{core}$ (M$_{\odot}$) \\
\hline
 & 0 & 9 & 4.37 & 3.59 & 0.7 & 0.28 & 3.4 \\
A & 33.04 & 9 & 4.2 & 4.09 & 0 & 0.98 & M$_\mathrm{He}$ = 1.5 \\
B & 33.04 & 15.7 & 4.08 & 4.65 & 0 & 0.98 & M$_\mathrm{He}$ = 1.5 \\
C & 33.07 & 15.7 & 4.41 & 4.61 & 0 & 0.98 & M$_\mathrm{He}$ = 1.5 \\
D & 36.55 & 15.7 & 4.19 & 4.62 & 0 & 0 & M$_\mathrm{CO}$ = 1.4 \\
\hline
\end{tabular}
\end{table*}

\begin{figure*}[t]
\centering
   \includegraphics[width=14cm]{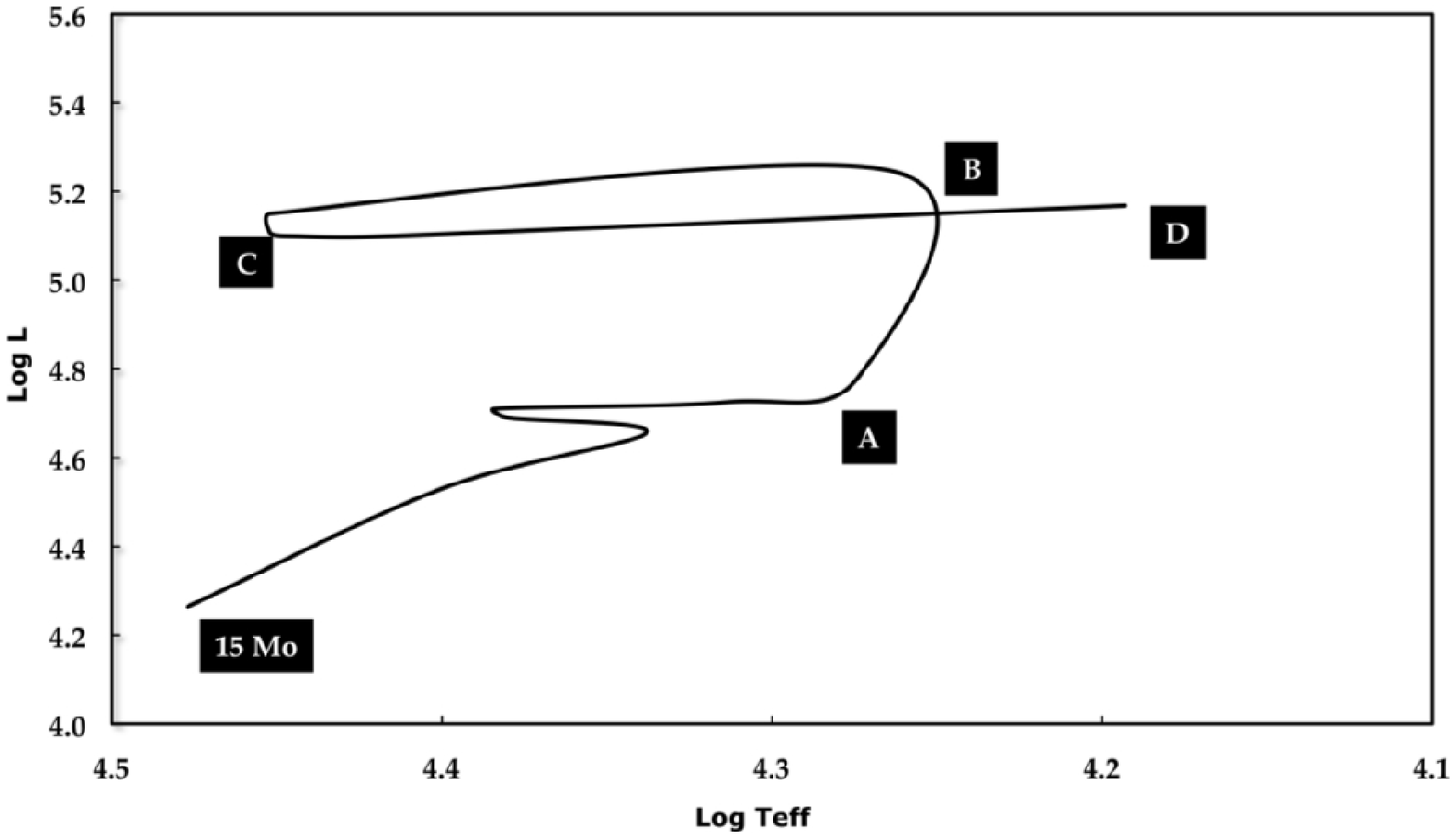}
     \caption{The evolutionary track of a 15 M$_{\odot}$ star in a Case Br binary that merges with its 9 M$_{\odot}$ companion. The labels have the same meaning as in Figure 1.}
\end{figure*}

\begin{figure*}[t]
\centering
   \includegraphics[width=14cm]{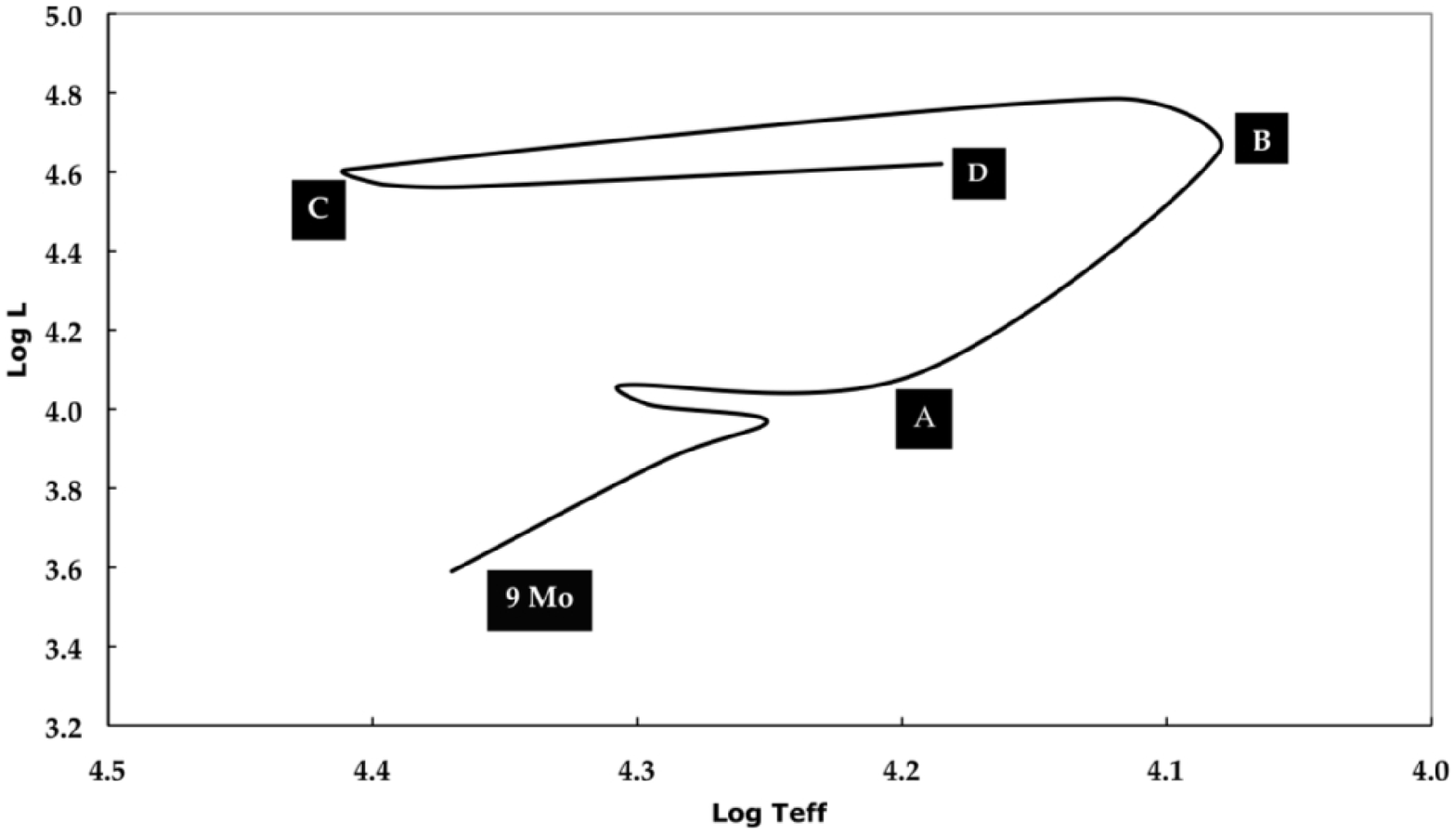}
     \caption{The evolutionary track of a 9 M$_{\odot}$ star in a Case Br binary that merges with its 6.7 M$_{\odot}$ companion. The labels have the same meaning as in Figure 1.}
\end{figure*}

We also evolved a 15 + 9 M$_{\odot}$ and 9 + 6.7 M$_{\odot}$ binary in the same way as the one discussed above. Figures 2 and 3 show the HRD tracks and tables 2 and 3 summarize evolutionary parameters. As can be noticed, the overall evolution is very similar and most interestingly, also here the post-merger hydrogen rich star remains blue during its entire core helium burning phase. 

A few remarks are appropriate.

1. It is highly probable that the merging process is accompanied by mass loss (Suzuki et al., 2007) but the amount is highly uncertain and we therefore did not account for such mass loss. Therefore our simulations apply to the case that this mass loss is much smaller than the mass of the secondary star that merges with the primary (which is likely to be the case when considering the results of Suzuki et al.). 

2. Our results obviously depend on the adopted model to mix the merging secondary with the outer layers of the primary. However, as long as the helium core of the merger is not affected, it can be expected that other mixing scenario's will lead to an evolutionary behaviour of the merger that is similar to the one discussed above.

3. Although the mass of the 30 + 20 M$_{\odot}$ merger falls in the mass range where single stars at the end of their evolution are expected to collapse as massive black holes, it has a helium core corresponding to a 30 M$_{\odot}$  star and the end of the merger will be marked by a supernova explosion in an extended hydrogen rich envelope with the formation of a neutron star. The magnetar model to explain SLSN-II (see also the introduction) may therefore fit into the present merger scenario.

4. As can be noticed from Figures 1, 2 and 3 the mergers spend most of their core helium burning as blue supergiants. Our models therefore predict that a significant fraction of the observed blue supergiants may be Case Br binary mergers (see also the next section). 

5. The simulations presented here are largely independent from the adopted initial composition in general, the metallicity Z in particular. The Z-dependence enters the simulations via the stellar wind mass loss formalism. In our stellar evolutionary code we use the De Jager et al. (1988) formalism as long as the star remains in the blue part of the HR-diagram and the surface hydrogen abundance X$_\mathrm{atm} > 0.3$. Massive stars with X$_\mathrm{atm} \le 0.3$ are defined as late type WN (WNL) stars for which we use WR-like stellar wind mass loss rates which are an order of magnitude larger than the De Jager rates. With this definition, our merger models never become WNL stars and therefore the effect of stellar wind mass loss (and thus of Z) on our simulations is rather modest. As a consequence, if our merger simulations reflect the nature of at least some SLSN-II then they predict the existence of SLSN-II also at Solar metallicity. Interestingly, at least 2 SLSN-II have been observed in galaxies similar to our Galaxy and this may indicate that SLSN-II also happen in regions with Z = 0.02.

Obviously, when the core helium burning mergers in our simulations have much higher stellar wind mass loss rates than those predicted by the De Jager formalism, independent from the value of X$_\mathrm{atm}$, then the effects of this mass loss on the core helium burning evolution of our mergers becomes more important as well as the possible Z-dependency of this mass loss. It is beyond the scope of the present paper to discuss this in more detail.

\section{The expected frequency of massive Case Br binary mergers.}

To calculate the frequency of massive Case Br binary mergers we use the Brussels binary population code as it has been described in Vanbeveren et al. (1998a, b) and in De Donder and Vanbeveren (2004). In particular, the results presented here are computed assuming that the initial mass function of the primary follows a Kroupa (1993)  slope, the binary mass ratio distribution is flat and the binary orbital period distribution is flat in the Log. In this case $\sim$48\% of the massive binaries are Case Br binaries. Sana et al. (2012) investigated binary properties of a statistically significant number of O-type stars in the Galaxy and in the Large Magellanic Cloud. They concluded that the binary frequency is large ($\ge 50$\%) and that the period distribution is skewed towards small periods. They propose a period distribution (Log P)$^{-0.55}$. Although it is unclear at present whether or not this applies for all the massive binaries (thus also for the early B-type binaries), we also made our simulations assuming that this period distribution applies for all massive binaries. Notice that in this case $\sim$37\% of all massive binaries are Case Br.

Obviously a detailed description of Case Br Roche lobe overflow is indispensable and is discussed below. The behaviour of a binary during Roche lobe overflow depends critically on the answer of the question: `what fraction of the mass lost by the primary due to the Roche lobe overflow process is accreted by the secondary (the parameter $\beta$) and, if mass leaves the binary, what is the process that makes this possible?' To our knowledge, in almost all detailed quasi-conservative Case Br massive binary evolutionary computations that have been performed by different research teams during the last 3 decades a situation is reached where during the rapid phase of the Roche lobe overflow both components reach a contact configuration. Although the contact may be a shallow contact, it is tempting to conclude that when this happens the further evolution is governed by a common envelope process where mass can leave the binary using the available orbital energy. If we know how much mass will leave the binary this way then the resulting orbital period evolution can be calculated using the $\alpha$-formalism by Webbink (1984) slightly modified to account for the fact that part of the mass lost by the primary due to Case Br Roche lobe overflow may be accreted by the secondary:

\[
\frac{M_{1i}\left(1-\beta\right)\left(M_{1i}-M_{1f}\right)}{\lambda R_{Roche}}
\]
\begin{equation}
\hspace{2cm}=\alpha\left(\frac{M_{1f}\left(M_{2i}+\beta\left(M_{1i}-M_{1f}\right)\right)}{2A_f}-\frac{M_{1i}M_{2i}}{2A_i}\right)
\end{equation}

\noindent where M$_1$ (resp. M$_2$) is the mass of the primary (resp. secondary), A is the binary separation, i and f stand for initial and final values, R$_\mathrm{Roche}$ is the Roche radius of the primary, $\lambda$ is determined by the density structure of the primary's outer atmosphere and on its internal energy that can help to expel the common envelope (see Dewi and Tauris 2000, for a detailed description and computation), and $\alpha$ is describing the efficiency of the energy conversion. Similarly as has been done in Vanbeveren et al. (2012) we will present population synthesis computations for $\alpha\lambda$ = 0.2, 0.5 and 1.

An attractive alternative is the model where matter escapes from the system via the second Lagrangian point $L_2$ from where it forms a circumbinary disk (van den Heuvel, 1993). To calculate the resulting period evolution we use the formalism presented by Podsiadlowski et al. (1992). In particular, this formalism contains the parameter $\eta$ that is defined as the diameter of the disk/orbital separation (obviously $\eta \ge 1$). Soberman et al. (1997) concluded that circumbinary disks are stable (e.g., the matter in the disk will not have the tendency to fall back towards the binary) only when their radii are at least a few times the binary separation and they propose $\eta = 2.3$. In the Brussels population code the latter is our standard (for comparison purposes, we will also present our simulations with $\eta = 1$).

Mass transfer during Roche lobe overflow is accompanied by angular momentum transfer and the mass gainer spins up. It was shown by Packet (1981) that a rigidly rotating main sequence star that accretes mass via a disk, reaches the critical rotation velocity when it has accreted roughly 5-10\% of its initial mass. One may be inclined to conclude that when this happens the remaining Roche lobe overflow has to proceed non-conservatively where mass can leave the binary via $L_2$ or via a common envelope like process by using the available orbital energy as discussed above. However, as shown by Paczynski (1991) and by Popham and Narayan (1991), mass transfer does not necessarily stop when the gainer rotates critically but how much mass can be further accreted is uncertain. We therefore made our simulations for $\beta$ = 0.1 and 0.5. Notice that the $\beta = 0.1$, $\eta = 2.3$ model for the 3 systems of the previous section corresponds with the common envelope model with $\alpha\lambda$ = 0.45 (30+20 M$_{\odot}$), 0.18 (15+9M$_{\odot}$) and 0.28 (9+6.7 M$_{\odot}$).

The criterion that we used to determine whether a given system will survive a mass transfer episode was to compare the theoretical stellar equilibrium radii of both stars after mass transfer (determined from their masses at that time) with the corresponding Roche radii\footnote{During the Roche lobe overflow phase neither the mass loser nor the gainer are in thermal equilibrium, however, when the mass loss/transfer phase stops, e.g., when the loser has lost most of its hydrogen-rich layers, both stars regain their equilibrium very rapidly.}. When at least one of the equilibrium radii is larger than the corresponding Roche radius, we conclude that the system merged. For systems that merge our population code calculates the matter lost through the non-conservative Roche lobe overflow up to that moment (see also Vanbeveren et al., 2012).

\begin{figure*}[b]
\centering
     \caption{The percentage (in gray scale given in the Legend) of Case Br mergers as a function of initial (ZAMS) primary mass and binary mass ratio for different combinations of $\beta$ and $\eta$ or $\alpha$. On top of each figure we also give the overall Case Br merger frequency. O+S means that the matter leaving the binary takes with the specific orbital and spin angular momentum of the gainer. In case the mass leaves the binary taking with only the specific orbital angular momentum (label O) of the gainer, there are no mergers. The percentages are calculated assuming a flat orbital period distribution. As outlined in the text this implies that $\sim$48\% of the massive binaries are Case Br.}\vspace{1cm}
     
   \includegraphics[width=14cm]{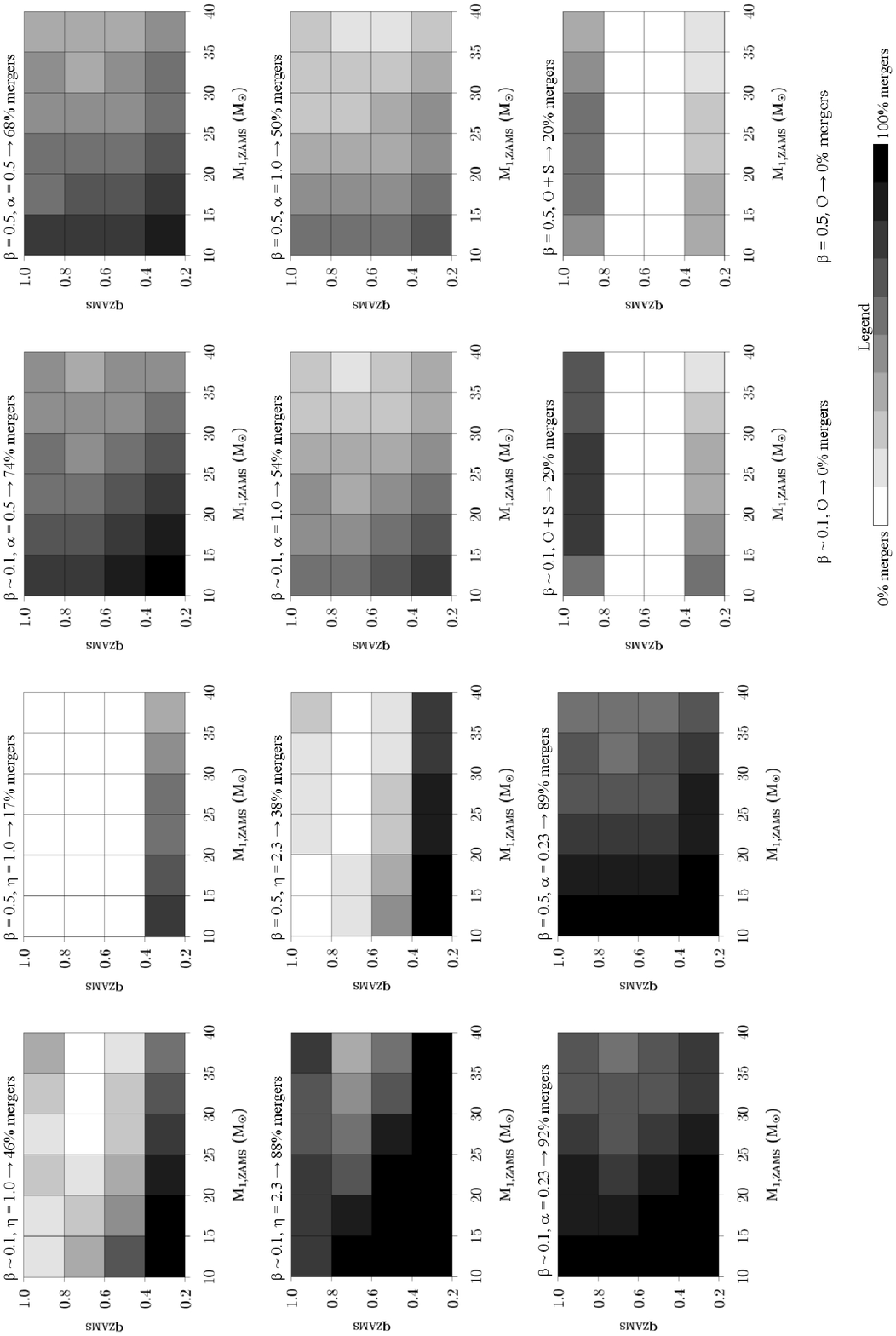}
   \end{figure*}
   
\begin{figure*}[b]
\centering
     \caption{Similar as Figure 4 but for a binary orbital period distribution $\sim$ (LogP)$^{-0.55}$. As outlined in the text this implies that $\sim$37\% of all massive binaries are Case Br.}
     \vspace{1cm}
   \includegraphics[width=14cm]{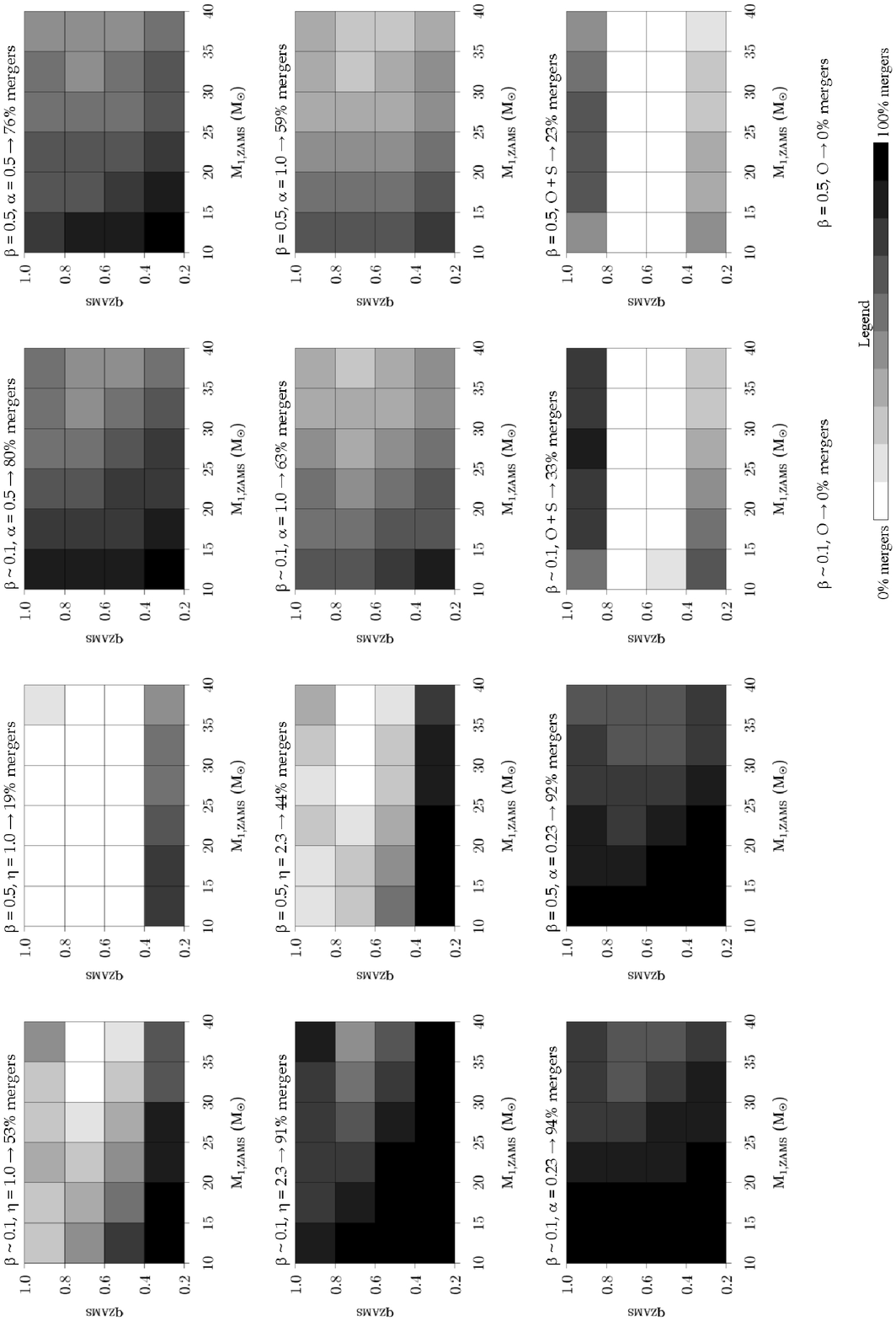}
   \end{figure*}

Figure 4 and 5 show the population of Case Br mergers as function of initial primary mass and binary mass ratio for different combinations of the parameters that govern Case Br evolution (Fig 4 is for the case of a flat Log P distribution, Fig 5 for the one proposed by Sana et al., see the discussion above). The figures illustrate that, if the Roche lobe overflow in Case Br binaries is non-conservative and if mass leaves the binary via a stable disk ($\eta = 2.3$) or via a process that is similar to the common envelope process, a significant fraction of the massive Case Br binaries merge. As expected, primarily the Case Br binaries with the largest periods survive.  In order to get an idea how many Case Br binaries may follow an evolutionary path comparable to the one discussed in the previous section, we also calculated the population of mergers where the mass of the mass gainer at the moment of merging is at least twice the mass lost by the loser and lost from the binary (e.g., the merger will have a He-core that is significantly undermassive with respect to the total mass of the merger, a necessary criterion for the merger to remain in the blue part of the HR-diagram). The results are shown in Figure 6. In combination with the expected frequency of massive Case Br binaries (see percentages given above) and with the fact that it is expected that these mergers remain blue for most of the core helium burning phase, it can readily be understood that our simulations predict that a significant fraction of the observed early B-type supergiants (perhaps also Be-type stars) are mergers of the type discussed here.

\begin{figure*}[b]
\centering
     \caption{The percentage of Case Br mergers where the merger has a helium core that is significantly undermassive (see text), thus for which the evolution discussed in section 3 applies. The results are calculated assuming a flat period distribution. The parameters and the legend are the same as in Figure 4.}
     \vspace{1cm}
   \includegraphics[width=14cm]{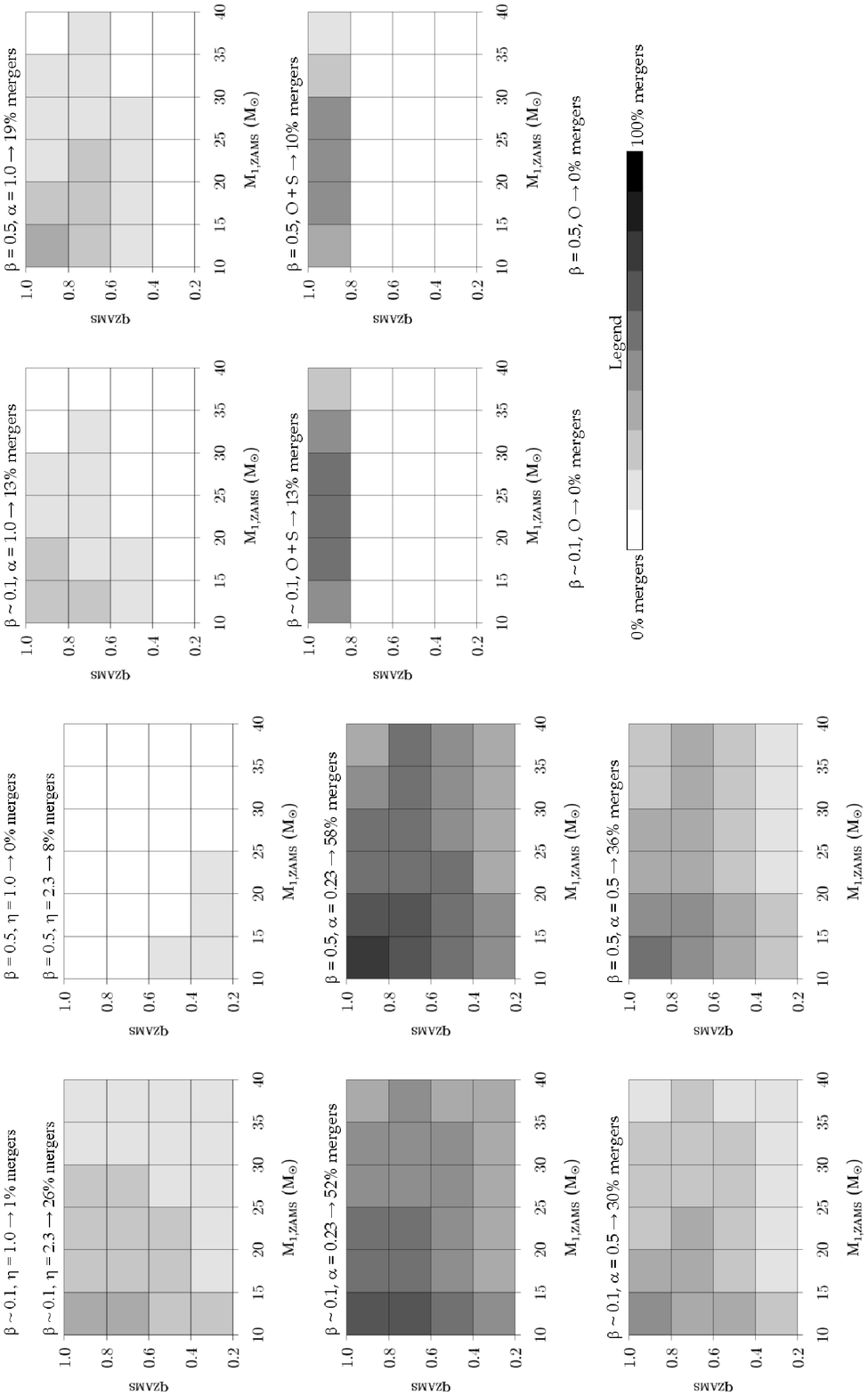}
   \end{figure*}

\paragraph{Remark}

The number of mergers resulting from massive Case Br binaries obviously depends critically on the adopted model that describes how mass will leave the binary. Compared to the $\eta = 2.3$ case the number of mergers is significantly smaller when $\eta = 1$. A model where it is assumed that mass escapes the binary via an enhanced wind from the gainer taking with the gainer's specific orbital angular momentum does not produce Case Br binary mergers at all. This model was adopted in many papers (e.g., Petrovic et al., 2005; De Mink et al., 2007 and many subsequent papers with authors of the same group). The assumption behind it is that when the gainer reaches critical rotation, its stellar wind is significantly enhanced such that the total amount of transferred mass leaves the binary by this wind taking with the gainers specific orbital angular momentum. However, rotation is not an efficient mass loss driver and the assumption is therefore highly questionable. Even more, if the gainer rotates very rapidly it can readily be checked that the specific spin angular momentum is at least as large as the orbital one. Figure 4 also gives the Case Br merger population when mass leaves the binary taking with it both, i.e., the specific orbital angular momentum and the specific equator spin angular momentum. Although the same criticism applies as given above (i.e., rotation is not an efficient mass loss driver), it can be concluded that also in this case a sizable fraction of the Case Br binaries merge. Mainly those with mass ratio between 0.8 and 1 will also meet the criterion for the binary scenario discussed in the previous section.

\section{Conclusions}

In the present paper we have shown that due to a non-conservative Roche lobe overflow in massive Case Br binaries (= binaries with a massive evolved hydrogen shell burning star and a main sequence companion) a significant fraction of them will merge. This merger may result in the formation of a massive single star with an undermassive helium core. This star then remains in the blue part of the HR-diagram during most of its core helium burning phase. Our models predict that many early B-type supergiants (and perhaps many early Be-type stars) may be Case Br type binary mergers. Furthermore, it is tempting to relate the most massive mergers to at least some of the superluminous supernovae SLSN-II where the engine is a magnetar.

      


\begin{thebibliography}{}

  \bibitem{1} Arnett, W.D., Bahcall, J.N., Kirshner, R.P., Woosley, S.E., 1989, ARA\&A 27, 629
  \bibitem{2} Braun, H., Langer, N., 1995, A\&A, 297, 483
  \bibitem{3} Claeys, J. S. W., de Mink, S.E., Pols, O.R., et al., 2011, A\&A, 528, 131
  \bibitem{4} De Donder, E., Vanbeveren, D., 2004, NewAR, 48, 861
  \bibitem{5} De Jager, C., Nieuwenhuijzen, H., van der Hucht, K.A., 1988, A\&AS, 72, 259 
  \bibitem{6} De Loore, C., Vanbeveren, D., 1992, A\&A, 260, 273
  \bibitem{7} De Mink, S.E., Pols, O.R., Hilditch, R.W., 2007, A\&A, 467, 1181
  \bibitem{8} Dewi, J.D.M. \& Tauris, T.M., 2000, A\&A, 360, 1043
  \bibitem{9} Fowler, W.A., Cauglan, G.R., Zimmerman, B.A., 1975, ARA\&A, 13, 69
  \bibitem{10} Gal-Yam, A., 2012, Science 337, 927
  \bibitem{11} Georgy, C., 2012, A\&A, 538, 8
  \bibitem{12} Heger, A., Woosley, S.E., 2002, ApJ, 567, 532
  \bibitem{13} Hellings, P., 1983, Ap\&SS, 96, 
  \bibitem{14} Hellings, P., 1984, Ap\&SS, 104, 83
  \bibitem{15} Iglesias, C.A., Rogers, F.J., Wilson, B.G., 1992, ApJ, 397, 717
  \bibitem{16} Kasen, D., Bildsten, L., 2010, ApJ, 717, 245
  \bibitem{17} Kippenhahn, R., Ruschenplatt, G., Thomas, H.-C., 1980, A\&A, 91, 175
  \bibitem{18} Knop, R., Aldering, G., Deustua, S., et al., 1999, IAU Circular 7128
  \bibitem{19} Kroupa, P., Tout, C. \& Gilmore, G. 1993, MNRAS, 262, 545
  \bibitem{20} Langer, N., 1991, A\&A, 252, 669
  \bibitem{21} Langer, N., 2012, ARA\&A, 50, 107
  \bibitem{22} Langer, N., Norman, C.A., de Koter, A., et al., 2007, A\&A, 475, L19
  \bibitem{23} MacFadyen, A.I., Woosley, S.E., 1999, ApJ, 524, 262
  \bibitem{24} Nugent, P., Aldering, G., Phillips, M.M., et al., 1999, IAU Circular 7133
  \bibitem{25} Packet, W., 1981, A\&A, 102, 17
  \bibitem{26} Paczynski, B., 1971, ARA\&A, 9, 183
  \bibitem{27} Paczynski, B., ApJ, 370, 597
  \bibitem{28} Pauldrach, A.W.A., Vanbeveren, D., Hoffmann, T.L., 2012, A\&A, 538, A75
  \bibitem{29} Petrovic, J., Langer, N., Yoon, S.-C., Heger, A., 2005, A\&A, 435, 247
  \bibitem{30} Podsiadlowski, P., Joss, P.C., Hsu, J.J.L., 1992, ApJ, 391, 246
  \bibitem{31} Podsiadlowski, P., Joss, P.C., Rappaport, S., 1990, A\&A, 227, 9
  \bibitem{32} Popham, R., Narayan, R., 1991, ApJ, 370, 604
  \bibitem{33} Sana, H., de Koter, A., de Mink, S.E., et al., 2013, A\&A, 550, 108
  \bibitem{34} Sander, A., Hamann, W.-R., Todt, H., 2012, A\&A, 540, 144
  \bibitem{35} Schaller, G., Schaerer, D., Meynet, G., Maeder, A., 1992, A\&AS, 96, 268
  \bibitem{36} Schwarzschild, M., Harm, R., 1958, ApJ, 128, 348
  \bibitem{37} Smith, N., Li, W., Foley, R.J., et al., 2007, ApJ, 666, 1116
  \bibitem{38} Smith, N., Chornock, R., Li, W., et al., 2008, ApJ, 686, 467
  \bibitem{39} Soberman, G., Phinney, E., van den Heuvel, E., 1997, ApJ, 327, 620
  \bibitem{40} Suzuki, T.K., Nakasato, N., Baumgardt, H., et al., 2007, ApJ, 668, 435 
  \bibitem{41} Umeda, H., Nomoto, K., 2005, ApJ, 619, 427
  \bibitem{42} Van den Heuvel, E.P.J., 1993, in 'Interacting Binaries', Saas-Fee Advance Course 22, eds. H. Nussbaumer \& A. Orr, Springer-Verlag: p. 263
  \bibitem{43} Vanbeveren, D., 1993, SSRv, 66, 327
  \bibitem{44} Vanbeveren, D., 2009, NewAR, 53, 27
  \bibitem{45} Vanbeveren, D., De Greve, J.P., de Loore, C., van Dessel, E.L., 1979, A\&A, 73, 19
  \bibitem{46} Vanbeveren, D., De Donder, E., van Bever, J., et al., 1998b, NewA, 3, 443
  \bibitem{47} Vanbeveren, D., Mennekens, N., De Greve, J.P., 2012, A\&A, 543, 4 
  \bibitem{48} Vanbeveren, D., Van Bever, J., Belkus, H., 2007, ApJ, 662, 107
  \bibitem{49} Vanbeveren, D., Van Rensbergen, W., De Loore, C., 1998a, A\&AR, 9, 63
  \bibitem{50} Vanbeveren, D., Van Rensbergen, W., De Loore, C., 1998b, The Brightest Binaries, (Dordrecht: Kluwer)
  \bibitem{51} Vink, J.S., de Koter, A., Lamers, H.J.G.L.M., A\&A, 362, 295
  \bibitem{52} Webbink, R., 1984, ApJ, 277, 355
  \bibitem{53} Woosley, S.E., 2010, ApJ, 719, L204
  \bibitem{54} Woosley, S.E., Blinnikov, S., Heger, A., 2007, Nature, 450, 390

\end{thebibliography}
\end{document}